\documentclass[cits]{PoS}
\usepackage{fontenc}
\usepackage{graphicx}
\usepackage{times}
\usepackage[small,it]{caption}
\usepackage{mathptmx}
\usepackage{amssymb}

\title{A statistical study of SDSS radio-emitters}

\author{\speaker{Mariangela Vitale}\thanks{International Max Planck Research School for Astronomy and Astrophysics at the Univerities of Bonn and Cologne, Auf dem H\"ugel 69, 53121 Bonn, Germany},$^a$ Jens Zuther,$^a$ Macarena Garc\'{i}a-Mar\'{i}n,$^a$ Andreas Eckart,$^a${$^b$} Marcus Bremer,$^a$ M\'{o}nica Valencia-S$^a$ and Anton Zensus$^b$\\
        \llap{$^a$}I. Physikalisches Institut, Universit\"at zu K\"oln, Z\"ulpicher Strasse 77, 50937 K\"oln, Germany\\
        \llap{$^b$}Max-Planck Instutut f\"ur Radioastronomie, Auf dem H\"ugel 69, 53121 Bonn, Germany\\
        E-mail: \email{vitale@ph1.uni-koeln.de}, \email{zuther@ph1.uni-koeln.de}, \email{maca@ph1.uni-koeln.de}, \email{eckart@ph1.uni-koeln.de}, \email{mbremer@ph1.uni-koeln.de}, \email{mvalencias@ph1.uni-koeln.de}, \email{azensus@mpifr-bonn.mpg.de}}

\abstract{The cross-correlation of the Sloan Digital Sky Survey Data Release 7 with the Faint Images of the Radio Sky at Twenty-Centimeters survey allows for a multiwavelength statistical study of radio-optical galaxy properties on a very large number of sources. The correlation we find between $L_{[20 cm]}/L_{[H\alpha]}$ and optical emission line ratios suggests that the origin of the emission in powerful radio-galaxies is of nuclear rather than of stellar origin. In particular, the spectroscopic classification in Seyferts, Low Ionization Narrow Emission Regions (LINERs) and star-forming galaxies provided by emission-line diagnostic diagrams peaks on the AGN region for higher $L_{[20 cm]}/L_{[H\alpha]}$ values. The trend differs from Seyferts to LINERs. The [NII]/H$\alpha$ vs. equivalent width of the H$\alpha$ line diagram confirms the LINER classification for most of those that have been identified with the traditional diagnostic diagrams. A small fraction of sources seem to be powered by post-AGB stars instead.
          }

\FullConference{Nuclei of Seyfert galaxies and QSOs - Central engine \& conditions of star formation\\
		November 6-8, 2012\\
		Max-Planck-Insitut f\"ur Radioastronomie (MPIfR), Bonn, Germany}

\ShortTitle{Origin of line emission in SDSS radio-galaxies}

\begin{document}

\section{Introduction}
Our knowledge of global galaxy properties has been recently improved thanks to large-area surveys such as the Sloan Digital Sky Survey (SDSS) \cite{York2000}. The cross-correlation of the SDSS Data Release 7 (DR7) with the Faint Images of the Radio Sky at Twenty-Centimeters (FIRST) \cite{Becker1995} survey offers the chance to study the optical properties of a large sample of radio emitters.\newline
The intensity of optical emission lines in radio AGNs has been already investigated in the past \cite{Saunders1989,Zirbel1995,Baum1995,Kauffmann2008}. The correlation between line strength and radio power suggests that optical and radio emission originate in the same physical process. However, the most powerful radio galaxies are generally detected at higher redshifts than the less powerful radio galaxies. This makes it difficult to establish whether there is a link between emission-line and radio luminosity, or between emission-line luminosity and redshift \cite{McCarthy1993}. The finding of a high number of detected AGNs with increasing redshift is also supported by the downsizing scenario of galaxy evolution \cite{Cowie1996,Thomas2005,Bundy2006}. According to this scenario, galaxies placed at higher redshift are more massive and host black holes that accrete producing powerful jets.\newline
AGNs can be selected from spectroscopic surveys using some optical emission line ratios. Low-ionization emission-line diagnostic diagrams \cite{Baldwin1981} are used to point out the connection between the galaxy nuclear activity, its morphological type \cite{Ho1997}, and its evolutionary stage \cite{Hopkins2006a}. We make use of optical and radio data to conduct a statistical study on the prospects of identifying radio galaxies in some well-defined regions of the diagnostic diagrams.

\section{Galaxy sample}
The Sloan Digital Sky Survey (SDSS) is a photometric and spectroscopic survey that covers one-quarter of the celestial sphere in the north Galactic cap \cite{York2000}. The spectra have an instrumental velocity resolution of $\sigma \sim 65$ km/s in the wavelength range $3800-9200~$\AA. The identified galaxies have a median redshift of $z\sim 0.1$. Spectra are taken with $3''$ diameter fibers ($5.7$ kpc at $z\sim 0.1$). We make use of the Max-Planck-Institute for Astrophysics (MPA)-Johns Hopkins University (JHU) DR7 of spectrum measurements (http://www.mpa-garching.mpg.de/SDSS/DR7/). The catalog contains the derived galaxy properties of the $\sim 10^6$ sources from the SDSS DR7 \cite{Abazajian2009}.\newline
The Faint Images of the Radio Sky at Twenty-Centimeters (FIRST) Survey \cite{Becker1995} makes use of the Very Large Array (VLA) in the B-array configuration to produce a map of the $20$ cm ($1.4$ GHz) sky emission with a beam size of $5''.4$ and an rms sensitivity of about $0.15$ mJy/beam. The survey covers an area which corresponds to the sky regions investigated by SDSS, and includes $\sim10^5$ sources. \newline
The crosscorrelation of SDSS with the FIRST survey provides a large optical-radio sample of galaxies. For generating the cross-matched FIRST/SDSS sample, we used the matching results provided by the SDSS DR7 via Casjobs \cite{OMullane2005}, based on a matching radius of $1''$. We apply the following conditions to our final sample: 1) $z>0.04$ to avoid aperture effect \cite{Kewley2006} and 2) relative error on the EW measurements of lines involved in the diagnostic diagrams $<30$\%. Our final cross-matched sample consists of $9\ 594$ objects, which correspond to $25.6$\% of the full cross-matched optical-radio sample, $9.6$\% of the original radio sample (FIRST) and $\sim1$\% of all the galaxies in the MPA-JHU data release.

\section{Optical emission-line diagnostic diagrams}
The Baldwin-Phillips-Terlevich (BPT) low-ionization diagnostic diagram \cite{Baldwin1981} and its subsequent versions \cite{Veilleux1987,Kewley2006,Lamareille2010} make use of emission line ratios. The strength of a line ratio is considered to be either a function of the hardness of the ionizing field of the galaxy and the metallicity \cite{Stasinska2008}. Higher ratios are assumed to mostly be the product of ionization that arises due to accretion around the black hole, rather than photoionization by hot massive OB stars. This diagnostic technique, largely used in the optical wavelength regime, allows differentiation of galaxies that show activity in their nuclei and starbursts. In particular, narrow-line AGN can be identified by the ratio of some distinctive emission-lines, such as [OI] $\lambda6300$ \AA\ over H$\alpha\lambda6563$ \AA\ and [OIII] $\lambda5007$ \AA\ over H$\beta\lambda4861$ \AA{}.\newline  
Transitions requiring relatively high ionization potential are found to happen in powerful radio emitters \cite{Baum1989,Morganti1992,Zirbel1995,Tadhunter1998,Best2012}, pointing to a possible correlation between AGN-detection rate and radio luminosity of the host galaxies. Some of the AGN and composite galaxies with widely spread distributions of metallicity and ionization parameters are degenerate with star-forming galaxies. This is especially the case of the [OI]/H$\alpha$ versus [OIII]/H$\beta$ diagram, which is particularly sensitive to shock excitation coming from either AGN or stars.
\begin{figure}
  \centering
  \includegraphics[width=10.5cm]{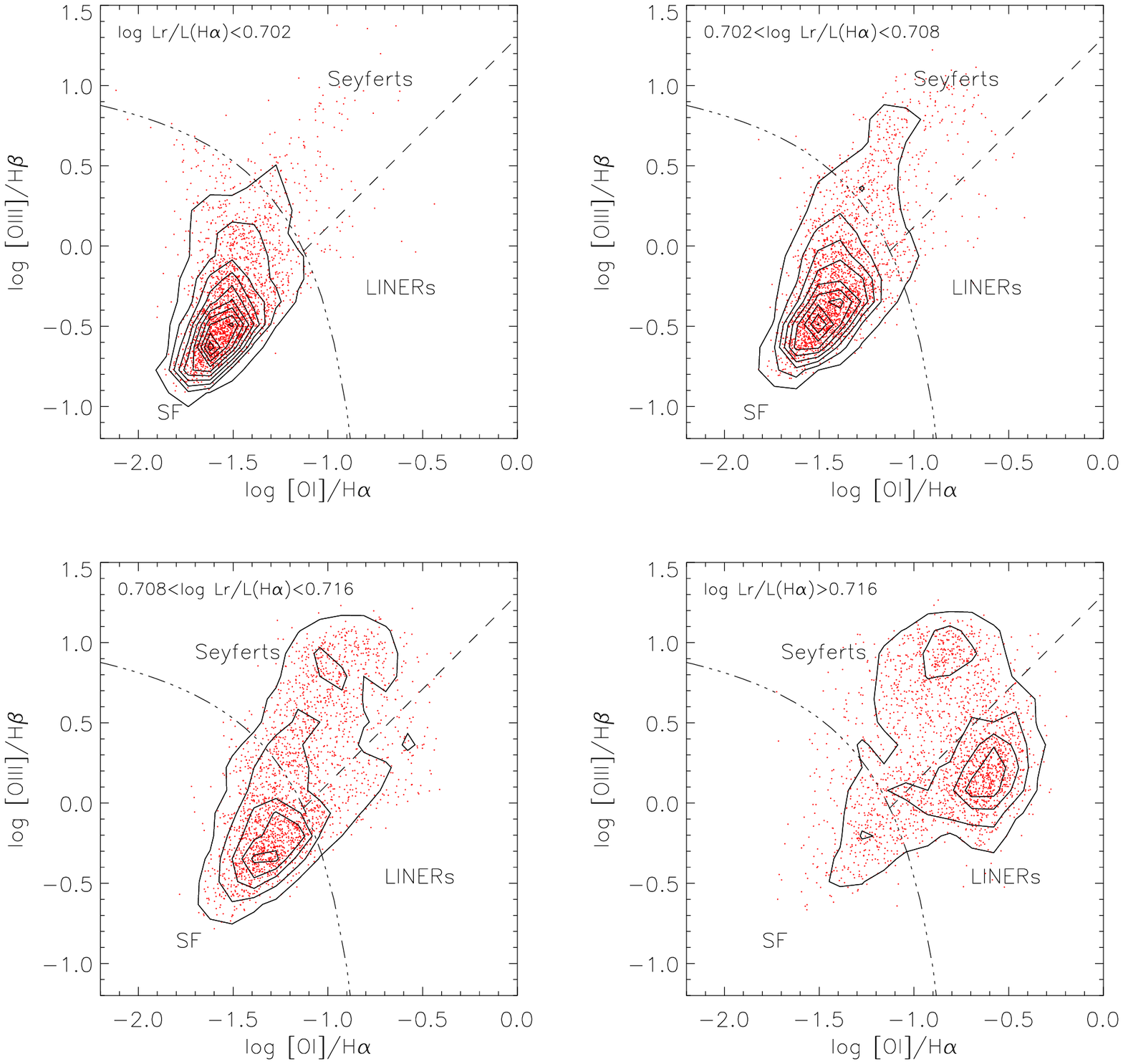}
  \caption{[OI]-based diagnostic diagrams. From top to bottom and from left to right, log$(L_{20~cm}/L_{H\alpha})$ increases. The number of radio emitters per bin is $2\ 350\pm25$. The contours represent the number density of the radio emitters ($20$ galaxies per density contour).}
 \label{oi_lum_tresh}
 \end{figure}

\section{Results}
\subsection{Trend with $L_{20~cm}/L_{H\alpha}$}
\begin{figure}
  \centering
  \includegraphics[width=12cm]{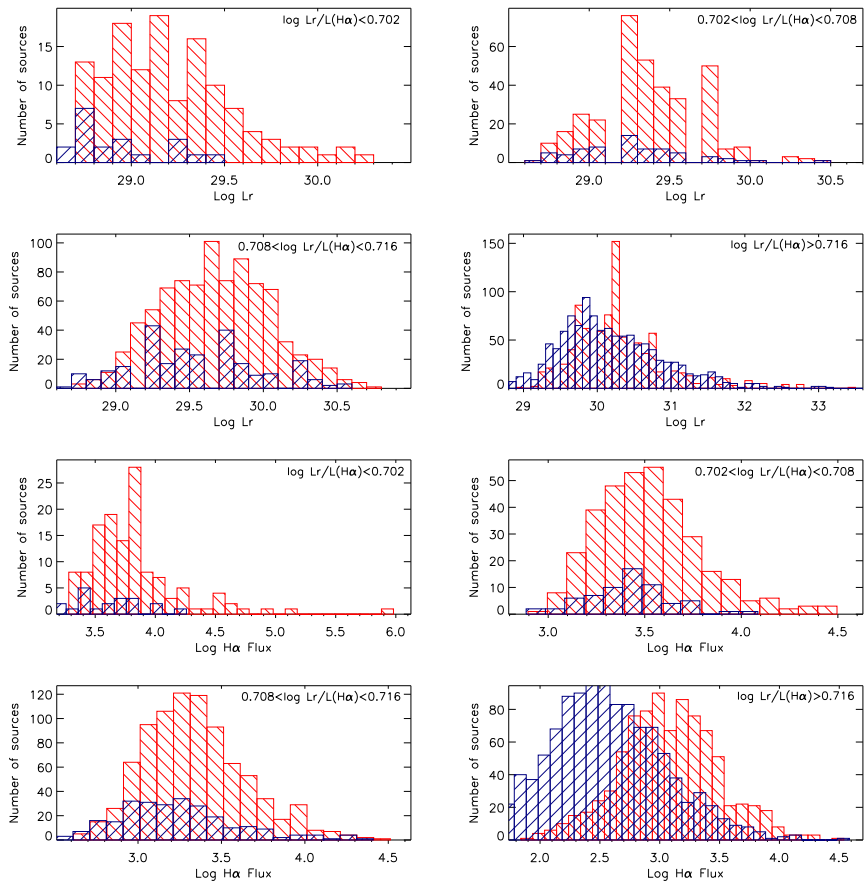}
  \caption{First and second rows: $L_{20~cm}$ distribution for Seyferts (red) and LINERs (blue) in the four $L_{20~cm}/L_{H\alpha}$ bins. Third and fourth rows: H${\alpha}$ flux distribution.}
 \label{histograms}
 \end{figure}
We study the evolution of the spectroscopic classification of galaxies of the optical-radio sample by placing them in four bins. The bins present increasing $L_{20~cm}/L_{H\alpha}$ values and contain approximately $2\ 350$ galaxies each. The binning has the purpose of searching for a threshold above which the objects are classified as AGNs (Seyferts or LINERs). The luminosity of the H$\alpha$ line, $L_{H\alpha}$, is considered to be a good optical star formation rate (SFR) indicator \cite{Moustakas2006}. The ratio between the radio luminosity, $L_{20~cm}$, and $L_{H\alpha}$ can be used to compare emission from radio components with the emission from young stars. The luminosity of the H$\alpha$ line has been derived after correcting the corresponding flux for the visual extinction, by using a theoretical H$\alpha/$H$\beta$ Balmer ratio of $2.86$. Our results are shown in Fig. \ref{oi_lum_tresh}. More diagnostic diagrams and a complete statistical analysis are offered in \cite{Vitale2012}. The curves separate the three spectroscopic classes of active galaxies. We find that the peak of the distribution shifts from the star-forming (SF) region of the diagrams to the composite (mixed contribution) or AGN part for increasing log$(L_{20~cm}/L_{H\alpha})$. In particular, the upper left-hand panels mostly show starbursts with high metallicity, while the bottom left-hand panels display a mixed population and the bottom right-hand panels show a nearly pure AGN population, together with some metal-rich starbursts. The distribution shows a peak in the LINER region for log$(L_{20~cm}/L_{H\alpha})>0.716$. In the upper right- and bottom left-hand panels ($0.702<$log$(L_{20~cm}/L_{H\alpha})<0.708$ and $0.708<$log$(L_{20~cm}/L_{H\alpha})<0.716$), the Seyfert region appears increasingly more populated, while the number of Seyfert galaxies remains constant for log$(L_{20~cm}/L_{H\alpha})>0.716$. In contrast, in the last bin we see an exponential increase in the number of LINERs.\newline
In Fig.\ref{histograms} (first and second rows) we show the $L_{20~cm}$ distribution for Seyferts (red) and LINERs (blue) in the four $L_{20~cm}/L_{H\alpha}$ bins. In the last three bins, the distributions span the same range of values. For log$(L_{20~cm}/L_{H\alpha})>0.716$, the distributions nearly overlap. In the third and fourth rows of Fig.\ref{histograms} the H${\alpha}$ flux distribution is shown. The last bin presents a clear dissimilarity in the distribution of Seyferts and LINERs, where the former peaks at higher values. This suggests that the exponential increases in the number of LINERs (Fig.\ref{oi_lum_tresh}, bottom-right panel) is probably due to the lower $L_{H\alpha}$ values that LINERs present compared to Seyferts.
\begin{figure}
  \centering
  \includegraphics[width=10.5cm]{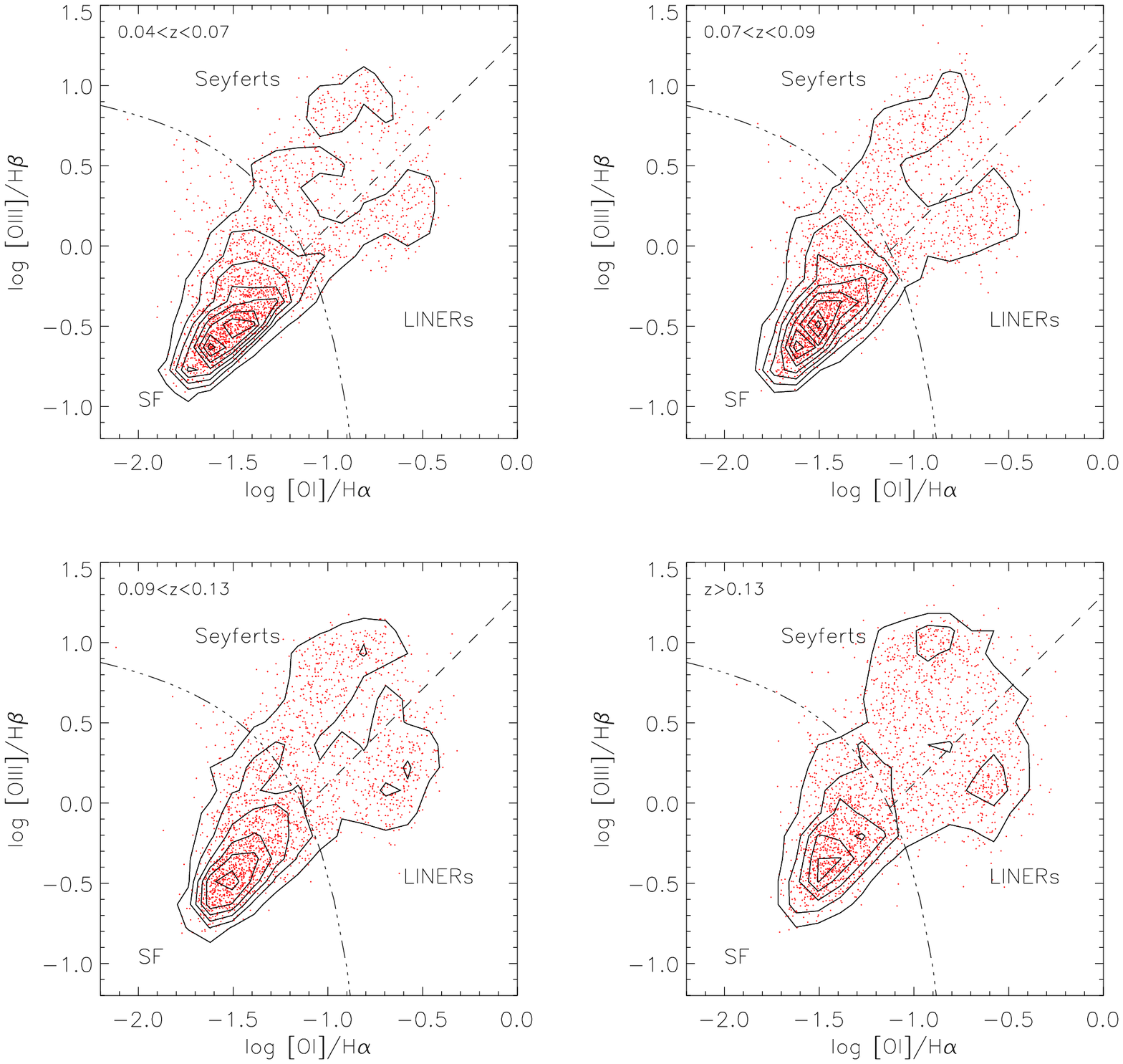}
  \caption{[OI]-based diagnostic diagrams. From top to bottom and from left to right, $z$ increases. The number of radio emitters per bin is constant and equal to $2\ 350\pm5$. The contours represent the number density of the radio emitters ($15$ galaxies per density contour). \label{oi_z_tresh}}  
 \end{figure}
 
\subsection{Trend with redshift}
We study the spectroscopic classification as a function of redshift by dividing the sample into four bins containing approximately $2\ 350$ objects each. We find that the distribution of radio emitters in the diagnostic diagrams depends mildly on the redshift (Fig. \ref{oi_z_tresh}). The number of AGNs always increases with $z$. The bulk of the population remains in the SF region in all $z$ bins. The increase in the number of LINERs and Seyferts show the very same trend. This result points to the possible interpretation that the higher number of AGN we select for increasing redshift is not only due to selection effects (in this case, we would select more powerful sources like Seyferts), but to a combination of the latter with true evolutionary effects.
\begin{figure} 
 \centering
  \includegraphics[width=10.5cm]{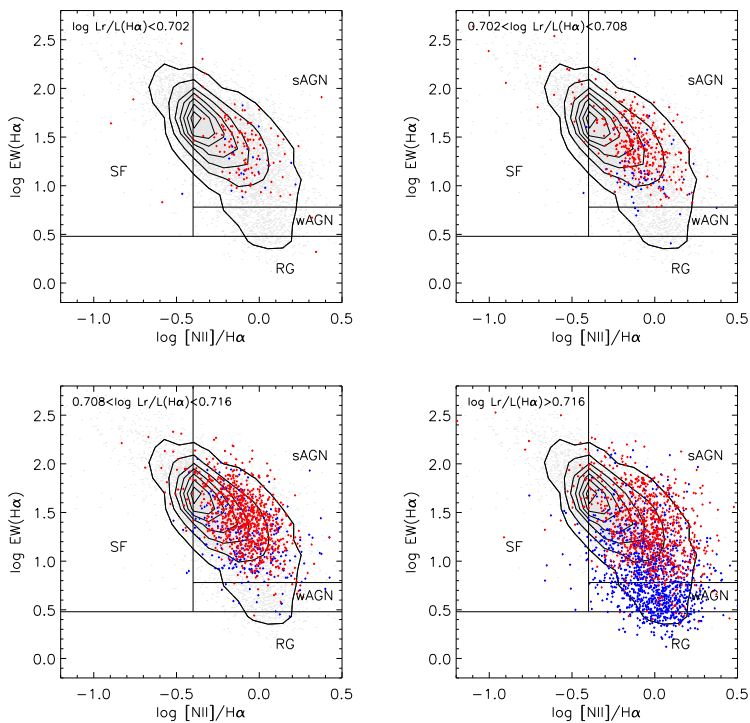}
  \caption{WHAN diagnostic diagrams for the optical-radio sample (light gray) overplotted with Seyferts (red) and LINERs (blue) selected from the [OI]-based diagram. Each plot represents, from the top-left to the bottom-right, a bin with increasing log$(L_{20~cm}/L_{H\alpha})$. Contours number density refers to the underlying light gray distribution of each plot and is equal to $70$ galaxies per contour.}
  \label{WHAN_L}
 \end{figure}

\subsection {Distinguish true from fake AGNs}
The question whether LINERs are true AGNs or not still needs to be answered. Emission from some of the galaxies classified as LINERs in the diagnostic diagrams is thought to be triggered by post-asymptotic giant branch stars and white dwarfs, which are abundant in early-type galaxies, classified as 'retired' (RGs) \cite{Stasinska2008}. Weak-line emitting galaxies have been found to have LINER-like emission with $3<$EW(H$\alpha)<6~$ \AA\ in the case of actual LINERs (labeled in the diagram as wAGN, where 'w' stands for 'weak', WHAN diagram \cite{CidFernandes2010}) and EW(H$\alpha)<3~$ \AA\ for retired galaxies. Galaxies that have both high values of EW(H$\alpha$) and [NII]/H$\alpha$ are classified as strong AGNs (sAGNs).\newline
Figure \ref{WHAN_L} shows WHAN diagnostic diagrams for the optical-radio sample, overplotted with Seyferts and LINERs selected from the [OI]-based diagram. Each plot represents a different $L_{20~cm}/L_{H\alpha}$ bin. While the distribution of Seyferts always peaks in the sAGN region, LINERs are mostly present in the sAGN region in the first three bins, and they appear to be mostly located in the wAGN region in the last $L_{20~cm}/L_{H\alpha}$ bin. The radio-strong (large $L_{20~cm}/L_{H\alpha}$) sources deviate from the RG and passive regions.

\section{Conclusions}
We have combined optical (SDSS DR7) and radio (FIRST) data to conduct a multiwavelength study on the evolution of physical properties of a large sample of radio emitters. We find that AGNs and radio galaxies with AGNs are drawn from a population that has higher metallicity than the overall SDSS sample. The lower abundance objects are predominantly star-forming, and populate the upper left-hand portions of these diagrams. \newline                                                                                                                                                                                                   
The connection between optical and radio properties is pointed out by the increasing number of Seyferts and LINERs that we find at high values of $L_{20~cm}/L_{H\alpha}$ and $z$. The spectroscopic classification is provided by optical emission-line diagnostic diagrams. The LINER population increases exponentially for the highest values of $L_{20~cm}/L_{H\alpha}$. LINERs in this luminosity bin present an increase in the radio power of the sources and the drastic drop in their $L_{H\alpha}$. The highest number of Seyferts are found in the third luminosity bin, where the sources present both a high $L_{20~cm}$ and $L_{H\alpha}$. The progressively and equally higher number of Seyferts and LINERs found at highest $z$ bins possibly indicate that the AGN-detection rate truly correlates with the redshift, besides being due to selection effects. The WHAN diagnostic diagram shows that the galaxies with LINER-like emission are mostly found to present the spectroscopic signatures of actual LINERs (assuming them to be low-luminosity AGNs).

{\scriptsize
 \bibliographystyle{abbrv}

}
\end{document}